\documentclass[conference]{IEEEtran}
\IEEEoverridecommandlockouts
\usepackage{cite}
\usepackage{amsmath,amssymb,amsfonts}
\usepackage{algorithmic}
\usepackage{graphicx}
\usepackage{textcomp}
\usepackage{xcolor}
\usepackage{url}
\usepackage[numbers]{natbib}
\usepackage{subcaption}
\usepackage{enumitem}
\def\BibTeX{{\rm B\kern-.05em{\sc i\kern-.025em b}\kern-.08em
    T\kern-.1667em\lower.7ex\hbox{E}\kern-.125emX}}
\begin{document}

\title{Low-cost Ultra-low Noise DAC System-on-Module \\for Scalable Ion-Trap Electrode Control\\

\thanks{This material is based upon work supported by the U.S. Department of Energy, Office of Science, National Quantum Information Science Research Centers, Quantum Systems Accelerator (Award No. DE-SCL0000121). Additional support is acknowledged from the National Science Foundation: Software-Tailored Architecture for Quantum Co-Design (STAQ) Award (PHY-2325080) and QLCI: Center for Robust Quantum Simulation (OMA-2120757).}
}
\author{\IEEEauthorblockN{1\textsuperscript{st} Mitchell G. Peaks}
\IEEEauthorblockA{\textit{Duke Quantum Center,}\\%, Department of Physics and Department of Electrical and Computer Engineering} \\
\textit{Department of Physics and}\\
\textit{Department of Electrical} \\
\textit{and Computer Engineering}\\
\textit{Duke University}\\
Durham NC, USA \\
mitchell.peaks@duke.edu}
\and
\IEEEauthorblockN{2\textsuperscript{nd} Mia M. Kaarls}
\IEEEauthorblockA{\textit{Duke Quantum Center and} \\
\textit{Department of Electrical}\\% and Department of Electrical and Computer Engineering} \\
\textit{and Computer Engineering} \\
\textit{Duke University}\\
Durham NC, USA \\
mia.kaarls@duke.edu}
\and
\IEEEauthorblockN{3\textsuperscript{rd} Crystal Noel}
\IEEEauthorblockA{\textit{Duke Quantum Center,}\\%, Department of Physics and Department of Electrical and Computer Engineering} \\
\textit{Department of Physics and}\\
\textit{Department of Electrical} \\
\textit{and Computer Engineering}\\
\textit{Duke University}\\
Durham NC, USA \\
crystal.noel@duke.edu
}}

\maketitle

\begin{abstract}
A new design for an open-hardware Digital-to-Analog Converter System-on-Module is presented for low-noise ion-trap electrode control. 
The design specifications were established to fill the technical needs of a modular, scalable DC electrode control platform with sufficient bandwidth, noise characteristics and control flexibility.  
Critically, a priority was placed on supply-chain management considerations and cost effectiveness for scaling. 
The system is based upon the Texas Instruments DAC81416 and AMD Xilinx Spartan-7 FPGA for the analog signal and compute architecture respectively.  
 Performance characterization of a prototype device suggests the design is suitable for a variety of ion-trap physics experiments and quantum computing applications.
\end{abstract}


\section{Introduction}
\IEEEPARstart{T}{rapped} ions are now a widely adopted platform for the study of quantum phenomena, as well as for development of quantum computers, simulators and sensors. Ion trap systems require a set of controllable DC electrodes to generate a static confining field in one principal trapping axis, for adjustment and trimming of the trapping potential, and to facilitate ion-transport operations \cite{Wineland1998ExperimentalIons,Jain2024, Delaney_2024}.  Increasing sophistication in the devices and architectural specifications have necessitated development of appropriate control electronics for these DC electrodes, not only in channel density, but also in noise characteristics and update bandwidth.
While there is often not a one-size-fit-all solution for DC electrode control, there are typically key features which are mutually desirable.  When considering digital-to-analog converter (DAC) Integrated Circuits (ICs), the specification trade-offs most relevant in this application are the update bandwidth, output noise, settling time, resolution, and output voltage range.  Power consumption is generally not a strong consideration due to the laboratory conditions not requiring portability or compact form-factors.  

An important set of factors not as typically discussed in smaller scale device production are supply chain issues, component sourcing, and cost effectiveness at scale.  
This type of analysis and engineering trade-off, while common in consumer markets and prototype electronics for production at scale, is rarely considered in experimental physics applications where many devices are one-offs or low production volume.  A notable exception to this may be ``big science" experiments such as those found at national or joint national facilities, for example in particle detectors or very large astronomical instruments \cite{Abi_2020_vol1, Abi_2020_vol3, McElwain_2023}.
These factors are becoming far more relevant as quantum technologies scale.  Dependence on specific products when assembling an experimental asset and then continuing to expand functionality can be difficult when component obsolescence or specific private entity development products are employed and are then no-longer supported.  
Given the lifespan of many of these experimental assets, and their continued evolution and expansion, this can be a serious problem often requiring extensive redesign. Over-reliance on legacy equipment can also hinder continued progress on broader scientific objectives.  For this reason, a strong focus in developing the instrumentation discussed herein is on providing an adaptable, expansible core, with an open-source hardware framework. We prioritize availability of components, open-source development tools, and a core architecture which can be built, expanded and adapted by the end-user from Original Equipment Manufacturer (OEM) components.

Among the current, most widely used DAC systems for ion-trap electrode control are the ARTIQ/Sinara products from M-Labs \cite{sebastien_bourdeauducq_2021_6619071}. 
% The hardware is modular and features a core FPGA based control unit (options available). Modules are connected via a Eurocard Extension Module (EEM) bus.  
For ion trap applications, the Zotino card is the standard choice, and employs the Analog Devices AD5372, 32-channel, 16-bit DAC quoting $210$~nV/$\sqrt{\text{Hz}}$ at $50$~kHz \cite{jordens_zotino_2019}.  For higher speed applications, the Fastino board is a higher-speed version of the Zotino based on the Analog Devices AD5542 ($60$~nV/$\sqrt{\text{Hz}}$ at $500$~kHz over a $1$~kHz bandwidth) \cite{jordens_fastino_2020}. The Fastino has channels that can be pushed to update at $2$~Msps simultaneously when using gateware acceleration \cite{fastino_wiki}.  Other developments in this field include attempts to increase the output channel voltage to accommodate faster and more robust shuttling operations in Quantum Charge-Coupled Device (QCCD) architectures for which the M-Labs ARTIQ/Sinara stack includes additional amplifier hardware expansions \cite{hv_amp_wiki}.  

While these commercial options are appropriate for many experiments, there are several key limitations addressed by the design presented here. First, the cost is prohibitive when considering an ion trap with one hundred or more electrodes (such as the standard Sandia National Laboratory Phoenix design \cite{Revelle2020-bq}). It is likely that the number of necessary DC electrode controls will only increase in future designs potentially greatly increasing the cost for a single system \cite{sterk2024multijunctionsurfaceiontrap}. Additionally, with a commercial product, the design is fixed and does not accommodate custom features. The hardware also has its own required ecosystem for integration. Finally, the Vanguard DAC aims for lower noise specifications than those provided by existing commercial options.

There are several examples of small-scale DAC development for laboratory environments. A noteworthy non-commercial example is the DAC system presented in reference \cite{Ohira_2025}. In this example, a hybrid approach has been taken with a development/evaluation board employed for the compute and control module portion (AMD Xilinx ZCU106) and a custom DAC ``Front-end'' built to provide high analog voltages to the trap.  
% The custom board is based on the Analog Devices AD9707 DAC and a set of  amplifier ICs to convert the current output of the DAC to a buffered high-voltage output, and boost the output voltage. 
% A ``latch-and-hold'' architecture is employed in this design to economize on channel density and provide for fast ion transport operations, implemented using a high-speed switch network.  
This hybrid approach is convenient for proof-of-principle experiments in academic laboratories and is not uncommon \cite{Lee_2021}. 
At the Duke Quantum Center (DQC), 
% for larger scale trapped-ion systems, micro-fabricated ion traps provided by Sandia National Laboratories have been widely employed \cite{Revelle2020-bq}.  These traps typically require control of many DC electrodes ($\approx$100 channels) for full functionality.  
% For several of these systems, 
a custom DAC solution has previously been favored for electrode control, owing to the high channel density, synchronicity, and real-time control requirements.  This system is based on a carrier board which provides twenty-five Quad High-Accuracy DACs (Texas Instruments TI DAC8734), on a single board with a socket to connect an XEM6010 evaluation board produced and discontinued by Opal Kelly based on the Spartan-6 FPGA.  Another example of a hybrid control approach, this system requires the use of external proprietary software to program and control, and an obsolete third-party product to function, making integration, development and future support challenging.
Attempts to solve channel density issues for scaling ion-trap systems have also been explored by employing in-vacuum DAC systems which can be externally configured, reducing the vacuum feed-through load and hardware constraints \cite{Stuart_2019}.  
For fast ion-transport operations on state-of-the-art micro-fabricated ion-traps, developments using RF hardware are being explored, such as the AMD Xilinx Zynq Ultrascale+ RF System-on-Chip, to provide the DC electrode voltages to $\approx$100 electrodes with high update bandwidth \cite{Dudley_2025}.  
The Vanguard DAC departs from these examples by eliminating the reliance on a third-party evaluation board.

In the following, the Vanguard DAC System-on-Module (SoM) prototype is presented. Section~\ref{sec:specs} begins with the requirement specification written for the device and the myriad design decisions.  Section~\ref{sec:design-implementation} outlines the Hardware and PCB design settled upon for realizing the specifications.  In Section~\ref{sec:digital_design}, the FPGA digital design developed for testing and minimum viable operation are discussed.  Characterization and measurements of the system are presented in Section~\ref{sec:characterization}, before a discussion of future improvements, modifications, continued development and open source contribution are presented in Section~\ref{sec:outlook}.

\begin{figure}[h]
\centering
\includegraphics[width=\linewidth]{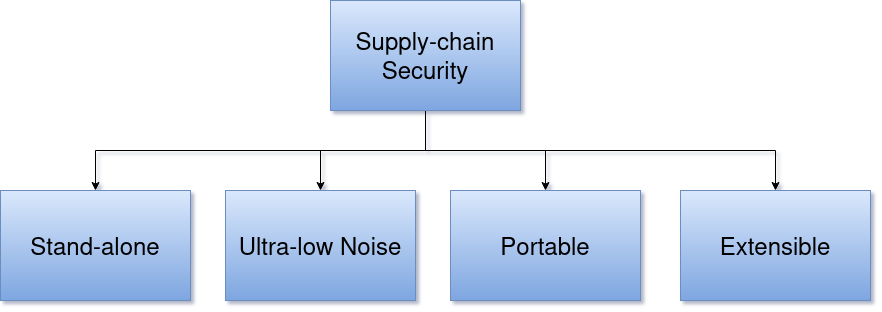}
\caption{System design requirements specification.}
\label{fig:requirements_spec}
\end{figure}

\section{Requirements Specification}
\label{sec:specs}
The following section lists the core requirements of the Vanguard DAC SoM. A summary of these requirements is depicted in Figure~\ref{fig:requirements_spec}. The requirements are written as a heuristic by which the device can be developed while maintaining and expanding the functionality and fitness for the core applications.  While the functionality might be expected to expand and/or branch throughout development for different or evolving system applications, the requirements specification provides a basis for current fundamental features to maintain in the core architecture. 

\subsection{Ultra-Low noise}

The primary specification for the MVP DAC SoM is to operate with sufficiently low-noise for trapped-ion devices.  
A lower bound on noise floor can be set by considering the projected Johnson noise from trap electrodes as the ideal noise-limiting feature of an ion-trap system. A typical ion-trap electrode might have a bulk resistance of $\approx$1~$\Omega$ leading to an equivalent spectral noise density of $0.13$~nV/$\sqrt{\text{Hz}}$ at $300$~K in a $1$~kHz bandwidth.  Practically, the Johnson noise floor is demonstrably well below the surface noise limitations in realistic ion trap systems \cite{Brownnutt2015Ion-trapSurfaces}.  This idealized noise floor does however provide a useful benchmark for design requirements in control hardware.
% As a result of this, a low-noise specification was required for the choice of DAC.
Given the requirements for scaling, supply-chain management, and control (discussed in the following), the noise specification must be weighed against cost per channel, availability, and voltage output specification etc. while remaining fit-for-purpose.
% In section \ref{sec:DACs}, we present the justification and device selection that was ultimately employed for the first version of the Vanguard DAC SoM (Texas Instruments DAC81416).

\subsection{Stand-alone operation}
 The device should be capable of operating as a stand-alone device without dependence on third-party control systems to operate.  This measure is applicable to the requirement for supply chain risk mitigation. However, we also ensure that electrical noise, bandwidth limitations, and rigidity of existing control architectures do not fundamentally limit the functionality of  this device.  The flexibility of the device should allow control development for existing, heterogeneous control systems e.g. ARTIQ (Advanced Real-Time Infrastructure for Quantum physics) \cite{sebastien_bourdeauducq_2021_6619071}, but it should not be fundamentally required.

\subsection{Supply chain security}
To ensure scalability and repeatability of the device, we generally require that components remain available and affordable. Specifically:
\begin{itemize}[itemsep=2pt, parsep=0pt, topsep=0pt, partopsep=0pt]
\setlength{\itemindent}{0in}
    \item All components should be available in stock from reputable vendors. 
    \item Fully supported and future supported components are reasonably assured i.e. no obsolete components and long-term support predicted. 
    \item Devices requiring programming e.g. FPGA, flash memory etc. programmable using free-to-use or open-source tools. 
    \item Design is maintained in-house with open-hardware architecture in entirety, thus removing dependence on private entity development/evaluation products. 
    \item Chosen components and development tools should have no known political or budget limitation issues. 
    \item Cost per channel should remain reasonable on component choice, particularly when considering the possibility of the requirements of extensibility and portability. 
\end{itemize}

\subsection{Extensibility}
Finally, we consider the aspects of the device which determine the ability to adapt the device for a specific use case. Basic logic elements and control hardware should be reusable, adaptable, and able to be expanded with deeper functionality or modified for different ion-trap control features such as fast-shuttling operations, real-time control, alternate pin-out or connector standards, or custom output filtering. 
% The design should be readily configurable to applications for which the basic design does not directly optimize  
Basic compute infrastructure should facilitate these changes without significant hardware modifications and allow soft reprogramming for different application and deployment.

\begin{figure*}[t!]
    \centering
    \begin{subfigure}[t]{0.5\textwidth}
        \centering
        \includegraphics[height=2.5in]{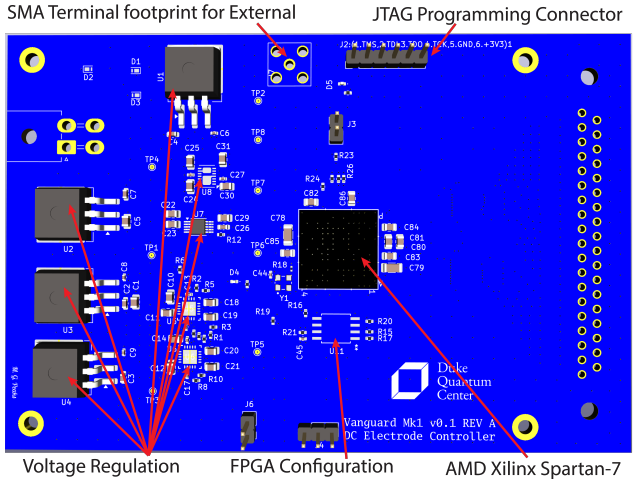}
        \caption{Front-view CAD model of the Vanguard System-on-Module hardware revision A.  This side houses the power supply components, digital logic and control hardware, centered on the AMD Xilinx Spartan-7 FPGA and supporting hardware.}
    \end{subfigure}%
    ~ 
    \begin{subfigure}[t]{0.5\textwidth}
        \centering
        \includegraphics[height=2.5in]{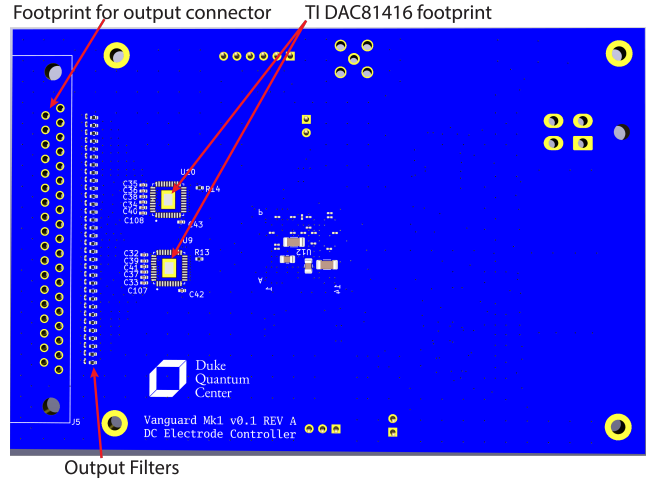}
        \caption{Rear-view CAD model of the Vanguard System-on-Module hardware revision A.  This is the analog side of the PCB, where the footprints for populating the DAC81416s can be seen with the RC filters to the output connector.  All other routing is kept to a minimum and physically distanced to avoid contaminating the noise environment of the DACs.}
    \end{subfigure}
    \caption{CAD model of the PCB design for v0.1 Revision A Mk1 Vanguard System-on-Module.}
    \label{fig:cad-pcb}
\end{figure*}

\section{Design Implementation}
\label{sec:design-implementation}
The hardware architecture was designed to fit the previously defined specifications for the device.  The following provides an explanation of the key components, design elements and their implementation.  A list of the key components (ICs) is provided in Table~\ref{tab:parts}, while a full bill-of-materials is included in the project directory (see \url{github repo}).  A model of the finalized Vanguard Mk1 v0.1 Rev. A DAC System-on-Module is shown in Figure~\ref{fig:cad-pcb}. 

\subsection{Power Supplies}
The inherent nature of the FPGA based logic and high-specification DACs for the application require multiple supply voltages, separate digital and analog signal domains, and a dual-rail supply to allow the DACs an appropriate output range at positive and negative output voltages.  Further, a robust and stable power supply solution should be compatible with low-noise laboratory linear power supplies.  For these reasons, only linear voltage regulators were employed in the design and any switched-mode voltage supply was deemed unfit. While linear regulators are inefficient and add a greater thermal load to the board than switching regulators, mitigating electrical noise is the highest-priority for the application. 
Two stages of voltage regulation are present on the board and serve several functions.  The voltage regulation stages step-down the typical range of supply voltages provided by the laboratory power supplies while distributing the thermal load, but also to act as a two-stage ripple-rejection filter on the input supply.
An additional benefit of this setup is that the more robust voltage regulators that are larger and functionally simpler to replace in the laboratory provide a layer of protection in the event of ESD or inadvertent over-voltage or surge current which may damage more vital components.  A set of TVS-diodes were also included on the input power supply stage to mitigate the effect of such events.

The second stage of power supplies was chosen for low-noise operation, minimal footprint, and to accommodate the specified logic power-up staging for the FPGA.  The digital systems supply was achieved with a pair of Texas Instruments TPS7A88 Dual low-noise, low drop-out (LDO) voltage regulators \cite{ti_tps7a88}, while the analog systems logic and DAC output supplies were achieved using the Analog Devices LT3042x \cite{adi_lt3042} and LT3032-12 \cite{adi_lt3032} respectively.

\subsection{Ground Plane}
Due to the mixed signal design and ultra-low noise requirement on the analog voltage control, the analog and digital power supplies are separated, and the PCB design minimizes cross-talk of the power planes.  A single ground reference across the board simplifies return paths and mitigates ground-loops, while minimizing loop areas for higher frequency signal paths both in the digital switching and analog DAC output portions.

%
%\begin{figure}[h]
%    \centering
%    \begin{subfigure}{0.1\textwidth}
%        \centering
%        \includegraphics[width=\linewidth]{images/main_power_input.png}
%        \caption{Caption for subfigure 1}
%        \label{fig:subim1}
%    \end{subfigure}%
%    \begin{subfigure}{0.5\textwidth}
%       \centering
%        \includegraphics[width=\linewidth]{images/second_stage_PSUs.png}
%        \caption{Caption for subfigure 2}
%       \label{fig:subim2}
%    \end{subfigure}
%    \caption{Caption for the main figure with two images}
%    \label{fig:main_figure}
%\end{figure}

\subsection{Compute \& Control Logic}
\label{sec:compute-control}
In order to allow for integration into various heterogeneous quantum control stacks, an FPGA was selected for the distributed compute and communication logic.  This choice makes the device highly adaptable, provides the ability to distribute control logic close to the quantum control system peripherals, and a platform upon which to build custom, low-latency gateware where required.  
The compute module remains adaptable, customizable, and expansible for different experimental systems.
% and provides a pathway for scaling.  
% The compute module was designed to be somewhat portable to alternative device layout and adaptation in other parts of the hardware, with the design being general enough to be employed across alternate or modified device architectures, while maintaining a 
It is also portable, in-house, and open-source to provide a pathway for scaling.  
The Xilinx Spartan-7 FPGA was chosen due to its availability, future support, and programmability with freely available tools.  The FPGA has sufficient resources to provide the embedded control logic while still being a low-cost option in the FPGA catalog.

Ion-trap DC electrode control typically requires precise voltage control with high stability. The update bandwidth must be commensurate with high-speed ion shuttling operations in the fastest use cases \cite{Bowler2012, Walther2012}.  
Strong low-pass filtering is often employed on DC ion-trap electrodes to mitigate noise \cite{Lee_2021}.  
This filtering typically provides the bottleneck to the analog voltage update bandwidth.  
The design decisions were thus taken to ensure that on-board control and communication logic should not bottleneck the update bandwidth.  Even the most modest specification for Spartan-7 architectures should be comparably negligible provided timing aware digital design is synthesized and implemented on the device (see section \ref{sec:digital_design}).  
In this case, the external clock provided on the board is a 50 MHz MEMS oscillator (Abracon ASEM1-50.000MHZ-LC-T), which provides a low jitter clock to a Multi-region Clock Compatible (MRCC) pin on the FPGA from which all synchronous operations are referenced.  
The hardware modules written for testing purposes (described in \ref{sec:digital_design}) are clocked clocked at 10 MHz using simple clock division to ensure a broad margin for communications timing setup and slack.  In this prototype, the low-pass filters on the DAC outputs ultimately limit the update rate (discussed in \ref{sec:DACs}).
% centralized or distributed systems that have synchronous operation across a networked control architecture, particularly as the control stack scales to larger systems.  
To provide a system which will be usable on a wide-range of experimental setups, the compute module remains a flexible prototype with a reconfigurable FPGA core that can accommodate developing control architectures that are increasing in scale.

\subsection{Digital-to-Analog Converters}
\label{sec:DACs}
Several candidate DAC ICs were vetted for use on the device, paying particular attention to the specified noise characteristics, settling time, resolution, cost-per-channel and domestic stock and availability, in line with the requirement specification.  Ultimately, the Texas Instruments DAC81416 was selected. 
The DAC81416 supports up to 16, 16-bit channels, and configurable bipolar voltage output ranges up to $\pm20$~V if required. The prototype SoM was designed with power supplies for $\pm10$~V operation, but later versions could be modified to support other ranges.  A specified output noise density of $78$~$\mu$V/Hz at $0.1-10$~Hz places the device within the ultra-low noise requirement.  The slew-rate is nominally 4~V/$\mu$s with a settling time of 12~$\mu$s, both of which provide a reasonably high potential update rate for ion-transport operations, comfortably above the pass-band of typical low-pass filtering in ion-trap DC electrode control schemata \cite{Akhtar2023}.  
The device supports communication via 4-wire serial interface up to 50 MHz which was employed for the versatile embedded logic control on the SoM \cite{ti_dac81416}.  For a $\pm10$~V configuration, 16-bit resolution corresponds to a LSB minimum voltage change between distinct codes of 305.176~$\mu$V.  The DAC81416 has an integrated 2.5~V precision internal voltage reference, with an initial accuracy of $\pm2.5$~mV maximum, and low-drift, order 5~ppm/$^o$C typical \cite{ti_dac81416}.  For this prototype SoM, the internal reference was used as a default mode however the device is also compatible with an external reference if access to a higher stability/precision voltage reference is available.

There are other DAC ICs available that provide even higher resolution and marginally lower output noise density.
One such example which was considered is the TI DAC11001A, which is available with $20$-bit resolution, $49$~nV/Hz output noise density and nominally 1~$\mu$s settling time.  The settling time is not a limiting factor, as previously discussed, as filtering typically provides the user-defined bottleneck. While there may be some merit to increasing DAC resolution for specific applications, and the output noise density is slightly lower, the cost-per-channel of this device is appreciably higher.  The DAC11001A being a single channel IC also greatly increases the physical footprint and routing complexity, with a single channel requiring a dedicated 48-TQFP land pattern on module $\approx$9~mm$^2$.  These factors provide a significant obstacle to cost-effectiveness, scaling, and accessibility, for relatively small benefits in terms of performance, which may not be relevant for the majority of applications dominated by other noise sources.
% wherein the noise and resolution characteristics are dominated by other sources in the broader ion-trap and control chain. 

\subsection{Filtering}
The connection of the DAC analog outputs are made to an external connector via low-pass filters.  These filters are topologically unipolar RC filters with a  designed $-3$~dB cut-off at $48.22$~kHz.  The inclusion of this specific filter was chosen to provide a limited high-frequency filtering but in a small form-factor and simple topology with a modest cut-off so as to maintain the flexibility of the base hardware design for multiple applications.  Custom external and in-vacuum electrical filters are typical in ion-trap systems, so the on-board filters are provided as a limited means of filtering to the output connector without over-specializing the DAC SoM.  

\begin{table*}[t]
\centering
\begin{tabular}{|p{0.08\textwidth}|p{0.12\textwidth}|p{0.18\textwidth}|p{0.41\textwidth}|}
\hline
\multicolumn{4}{|c|}{\textbf{Key Components}} \\ 
\hline \hline
\textbf{Designator} & \textbf{Manufacturer} & \textbf{Part No.} & \textbf{Description} \\
\hline
U1  & ST            & L7815             & Linear Voltage Regulator 1.5A 15V \\
U2  & Onsemi        & MC7915ACD2TG      & Linear Voltage Regulator 1.5A -15V  \\
U3  & ST            & L7806             & Linear Voltage Regulator 1.5A 6V \\
U4  & ST            & L7805             & Linear Voltage Regulator 1.5A 5V \\
U5  & Analog Devices & LT3042xMSE       & Ultra-Low Noise, Ultra-high PSSR Linear Regulator \\
U6  & Analog Devices & LT3032-12        & Dual 12V LDO Positive/Negative Low-Noise  \\
U7  & Infineon      & S25FL128L         & QSPI 133 MHz, 128-Mbit Flash Memory for FPGA configuration \\
U8  & AMD Xilinx    & XC7S6-1FTGB196C   & FPGA \\
U9  & Texas Instruments & TPS7A8801RTJR  & Dual Low-noise LDO \\
U10 & Texas Instruments & DAC81416RHAT   & 16-channel, 16-bit Digital-to-Analog Converter, Ultra-low Noise \\
U11 & Abracon       & ASEM1-50.000MHZ-LC-T & 50 MHz MEMS Oscillator to provide on-board clock  \\
\hline
\end{tabular}
\captionsetup{width=0.79\textwidth}
\caption{List of key active components used in the design.  The designators are references given in the PCB design files, while part numbers provided are OEM specified.}
\label{tab:parts}
\end{table*}

\subsection{PCB layout and stack-up}
\label{sec:PCB-layout}
\begin{figure}[h]
\centering
\includegraphics[width=\linewidth]{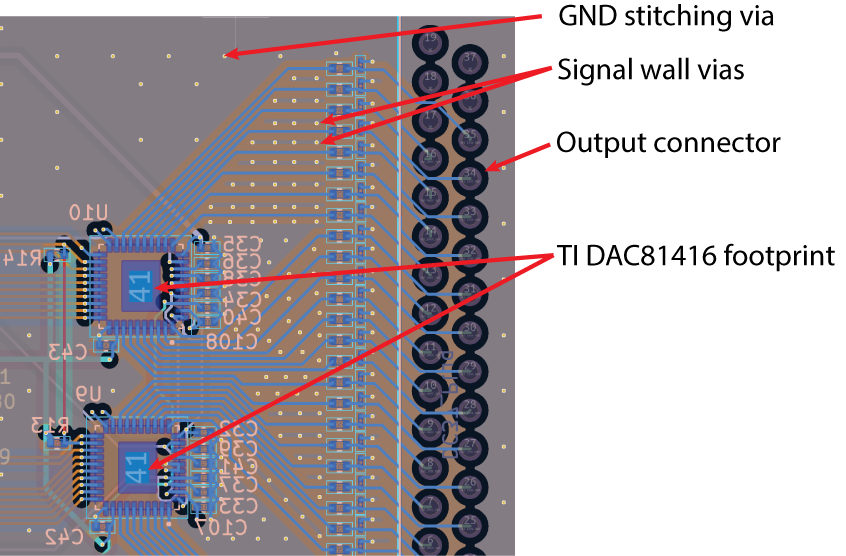}
\caption{Close-up of the PCB design showing the TI DAC81416 footprints, DAC output traces connecting to the external connector pads, and the EMI wall and stitching vias placed across the board to prevent EMI, channel cross-talk and to stitch the PCB ground planes.}
\label{fig:DACs_PCB_close}
\end{figure}

For the PCB layout and stack-up, multiple design decisions were made based on the specification, with particular care and attention to ensuring good isolation of digital and analog signal pathways.  An 8-layer PCB stack-up was employed as a means of providing dedicated copper planes for digital and analog power and signaling, each with a nearby ground plane.  The layout is designed to minimize inductance and signal return loop area to reduce Electromagnetic Interference (EMI) and noise coupling.  Extensive through and transfer vias were employed in and around the signal traces also to mitigate EMI, cross-talk and for thorough stitching of the ground planes to maintain an equipotential ground reference and prevent ground loops.  An example of wall-vias can be seen in Figure~\ref{fig:DACs_PCB_close} between DAC channel connections to the external connector pins, with the ground-plane stitching via matrix also visible across the board.

\section{FPGA Digital Design}
\label{sec:digital_design}

For the prototype, a minimum viable control system was built in FPGA hardware defined using Verilog HDL.  The control flow is a simple and effective system of defining the desired voltage outputs and DAC system behavior on a host computer, transmitting the required data to the FPGA DAC SoM, and allowing the DAC to be programmed accordingly by the control logic built on the FPGA.  This methodology allows for synchronous, real-time control via uploading the data asynchronously, and storing it on FPGA memory.  The device is then sent a digital logic trigger to start the pre-loaded process via a separate control line designed into the hardware (wired to an I/O pin on the FPGA).  

This methodology circumvents the complexity of direct real-time communication, which can be particularly problematic across a modular, scalable control architecture, often requiring a high bandwidth real-time communications protocol and hardware for synchronous transmission of control data.  This compute hardware setup, as described in \ref{sec:compute-control}, provides flexibility to the user for the myriad real-time applications in the quantum control architecture application space, e.g. in-circuit ion-transport \cite{Delaney_2024}, active ion micro-motion compensation etc. \cite{Berkeland_1998,Hogle_2024}.
The onboard compute module provides a system by which control logic can be distributed to the peripherals in the control stack, mitigating the need for time consuming communication back to a central control node, useful for high bandwidth real-time operation in larger scale systems.  

It is noteworthy however, that high bandwidth real-time communication systems, while adding complexity, may become necessary as modular control systems scale to very large channel number and to synchronize complex operations with other control signals required in the quantum computing stack e.g. high-frequency RF waveform generation.  Given the flexibility and open nature of the hardware design, later iterations accommodating high bandwidth real-time communication between boards are possible with continuing hardware development.

\begin{figure}[h]
\centering
\includegraphics[width=\linewidth]{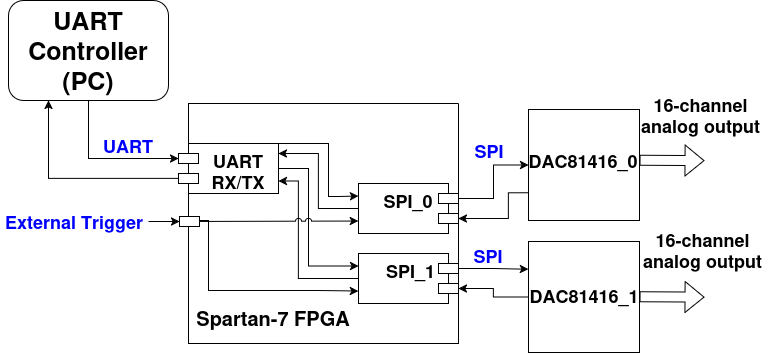}
\caption{Diagrammatic representation of the minimal control infrastructure build for testing the SoM.  The communication protocols are delineated in blue, and the hardware modules which handle communication are implemented in FPGA fabric.}
\label{fig:control_diagram}
\end{figure}
\vskip8mm
The minimal control build was implemented as follows: A python script runs on a local computer to communicate the 16-bit voltage value code, register address of the DAC channel on which the voltage is to be set, and the DAC device (i.e. DAC\_0 or DAC\_1 on the board).  The script puts the data on the COM port and a USB-to-serial interface cable transmits the control data to the SoM via UART.  A UART module, written in Verilog HDL and implemented on FPGA fabric, receives the voltage code, DAC index and DAC channel register address, and passes this, conditionally, to the relevant SPI controller module, also implemented on FPGA fabric.  The SPI controllers then communicate the data to the DACs, while also providing the necessary register instructions to activate and configure the DACs according to the desired output state.  This configuration register addressing includes powering on all channels, setting the voltage output range, configuring single or differential output, choosing to use the internal voltage reference etc.  A diagrammatic representation of the minimal control architecture is provided in figure \ref{fig:control_diagram}.

\section{Device Characterization}
\label{sec:characterization}
In order to benchmark operating characteristics of the prototype SoM, measurements were taken to evaluate the performance of the DAC output in-system, running with the aforementioned control logic (described in \ref{sec:digital_design}).  The measurements give an indication of the current fitness-for-purpose and can be used as a comparison to the isolated DAC specification provided in the datasheet.  It should be noted that changing the FPGA logic, including the clock frequency and distribution, specific synthesis, implementation and complexity of the gateware may influence the measured output characteristics, but the board is designed where possible to minimize the effect these changes may have on the analog system behavior.

\subsection{Output Voltage}
\label{sec:measurement_1}
We first aim to characterize the output voltage reproduction from register writes and output voltage stability.  A set of measurements were taken across a set of channels from each DAC on the SoM sampled from each register that is independently initialized.  The measurements were taken for a set of voltage outputs that are representative of relevant fixed control voltages.  The data was taken using a Keithley 2002 high-precision multimeter with $28$-bit ADC resolution corresponding to $100$~nV at $20$V and $1$~nV at $200$~mV range.  The measurements were taken by connecting each output on the board individually to the multimeter probe and referencing to the SoM ground-plane via a plated through-hole on the PCB.  Each channel was measured with an output corresponding to a 16-bit word register write, which defines the output voltage as a 16-bit division of the configured voltage range, in this case -$10$~V to +$10$~V with $305.176$~$\mu$V increments.    It should be noted that due to fixed point division when mapping the $20$~V range to $2^{16}$ divisions, the voltage division mapping is not strictly linear, e.g. while writing FFFF$h$ to the DAC output register yields an ideal voltage of $10.000000000$~V, writing 7FFF$h$ yields $-0.000152590$~V, not $0.000000000$~V.
 
\begin{figure}[h]
\centering\includegraphics[width=\linewidth]{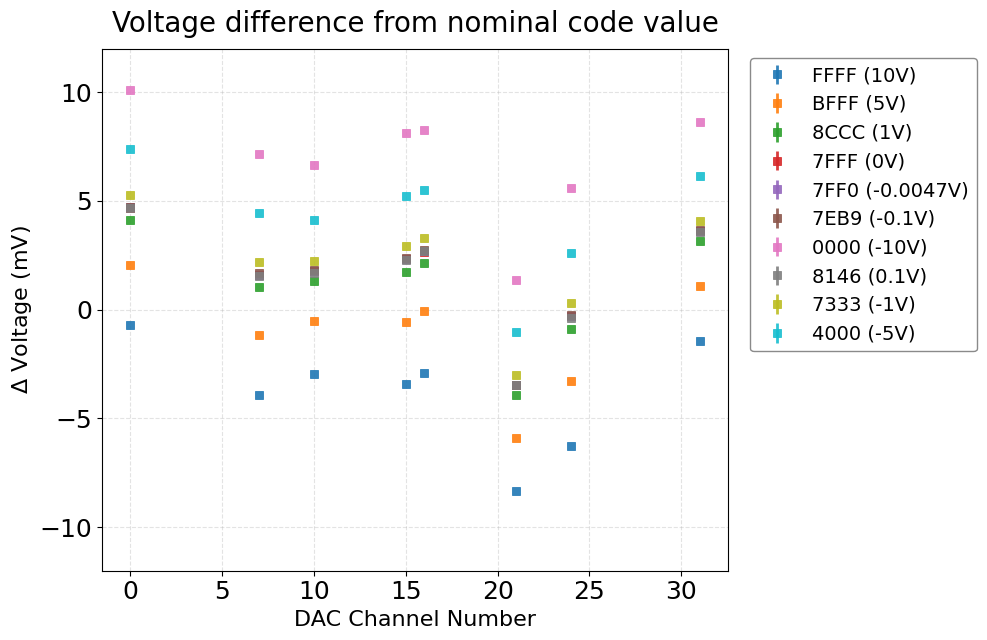}
\caption{Measurements taken at a sample of the output channels (samples taken from each power register) for a set of 16-bit word register write values corresponding to a 16-bit division of the DAC output range (-10~V to +10~V). $\Delta$Voltage represents the difference between the target value in the legend and the measured value.  The legend gives rounded values of the nominal code voltage correspondence for reference purposes.}
\label{fig:output_measurements}
\end{figure}

The data presented in Figure~\ref{fig:output_measurements} show the output voltage measured on channels sampled from each of the registers from which they are powered and controlled, in banks of four.  While there is minor channel variation, the output falls within the specified offset error given in the datasheet for the DAC, and is consistent and reproducible.  Error bars for this voltage measurement are asymmetric and provided based on the uncertainty in measurement from fast fluctuation on the order of $\pm30$~$\mu$V. A maximum and minimum measurement was taken using the multimeter in a $60$~s measurement window, to guarantee a fixed minimum and maximum value.  This window was determined to be sufficient by leaving the voltage measurement for $2$~hours on a small sample of channels, and verifying the maximum and minimum values were identical to that measured $30$~s after setting the respective register value.

\subsection{DAC Resolution and Offset}

\begin{figure}[h]
\centering\includegraphics[width=\linewidth]{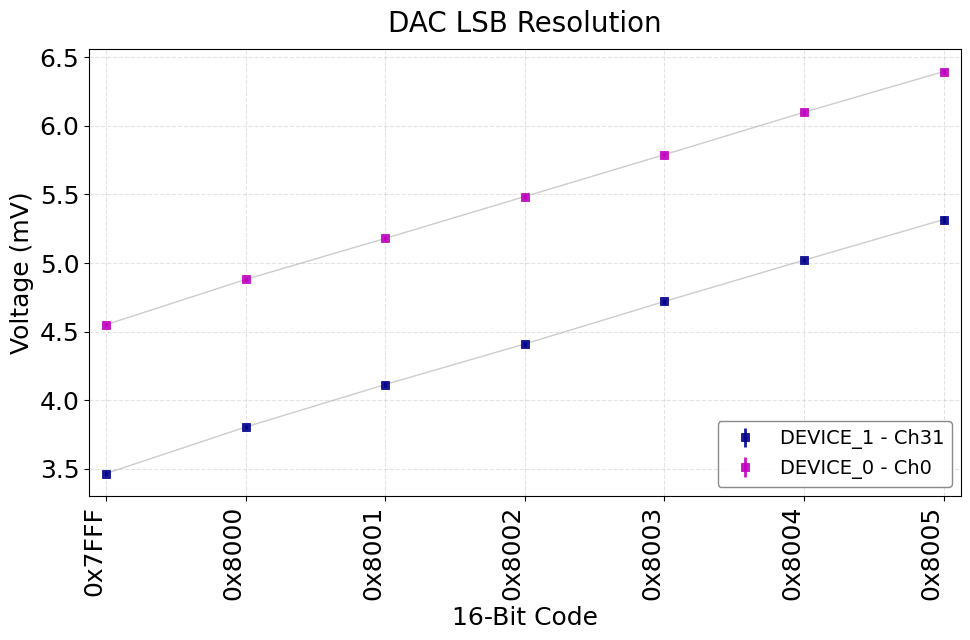}
\caption{Output voltage measured on on each of the two DACs on module at register write values in the center of the output range near 0~V (7FFF$h$ - 8005$h$). The value for each channel at 7FFF$h$ indicates the offset for that DAC from nominal 0~V.}
\label{fig:resolution_plot}
\end{figure}

The resolution and offset between the two DACs was observed by applying a set of output values near zero volts to a channel on each of the devices. Figure \ref{fig:resolution_plot} shows the least significant bit (LSB) incremented in the DAC output register on each of the two devices on the SoM.  The average LSB increment was measured to be $308(32)$~$\mu$V, which corresponds to the expected output change at this output range calculated as $305$~$\mu$V, accounting for measurement uncertainty and the specified differential non-linearity of $\pm 0.5$~LSB \cite{ti_dac81416}.  An offset of $1.07(3)$~mV was measured between the two devices.  This could be attributed to systematic effects in the test setup such as local charging or connectivity of the measurement probes, but may also be attributed to small variations in DAC device fabrication and the internal reference voltage.  Future iterations of the hardware could be modified to use a common external voltage reference or to have both DACs referenced to the internal reference of one of the devices to attempt to minimize voltage offsets between devices.  These static offsets can however be accounted for during calibration when deployed in application.

\subsection{Output Noise Spectrum}

\begin{figure}[h]
\centering
\includegraphics[width=\linewidth]{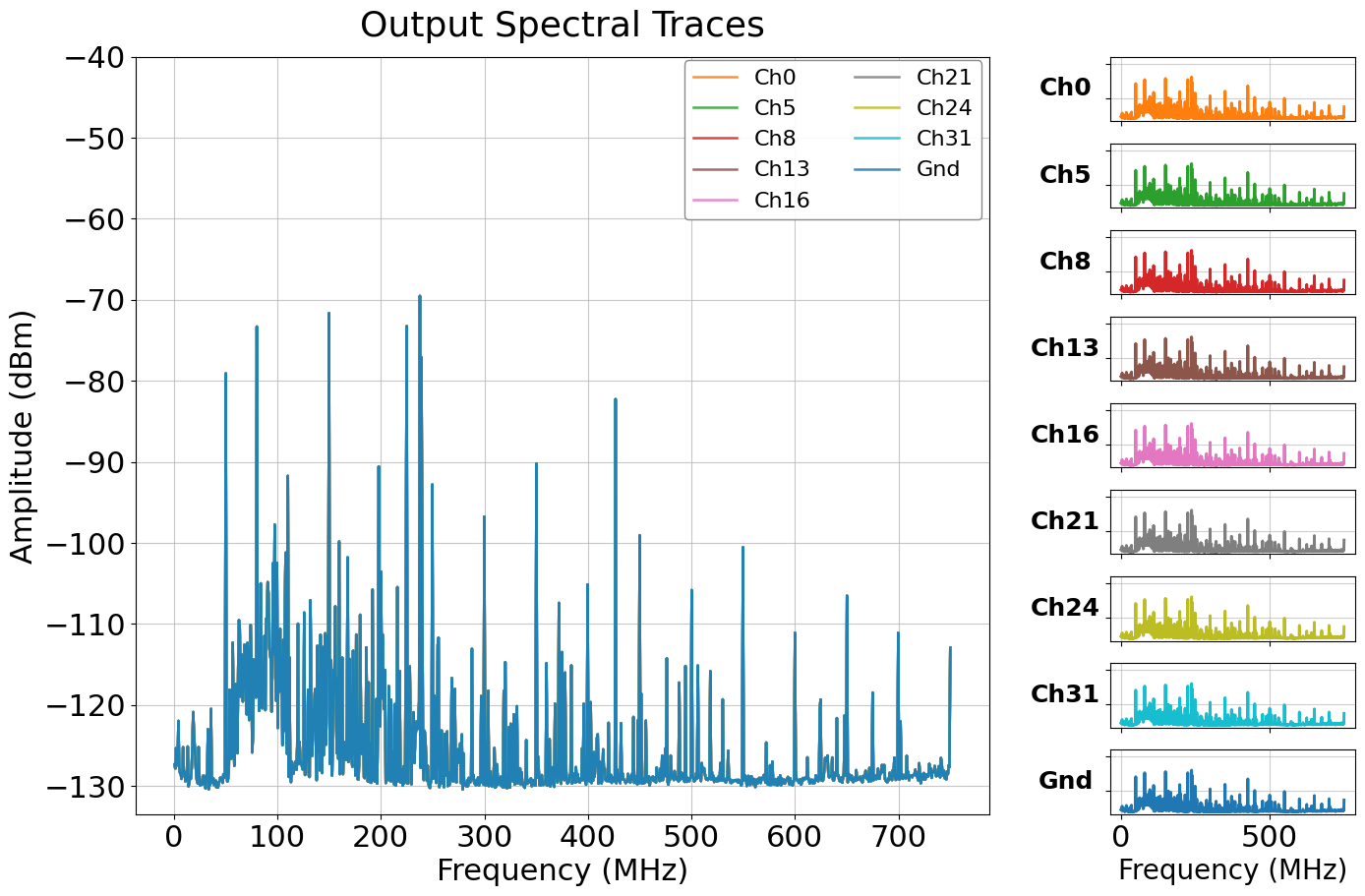}
\caption{Power Spectral Density trace taken in $2$~Hz to $750$~MHz band at $10$~Hz resolution bandwidth for a set of output channels corresponding to each power up/down control register (arranged in banks of four).  The main axes show an overlay of all traces where only a single trace is visible on account of the spectral indistinguishability.  The right column is representative of the individually plotted traces for each channel.}
\label{fig:spectrum}
\end{figure}

To determine whether additional noise appears on the DAC channels while operating the SoM, a spectrum analyzer was used to monitor the frequency spectrum on the output operating at a set of voltage outputs. A DC block was employed to filter the DC voltage output and protect the measurement instrument.  A baseline ground measurement was established by making a measurement with the probe on the ground plane of the SoM taken self-referentially, labeled Gnd in Figure~\ref{fig:spectrum}.  The following measurements were taken with the voltage output set to $-10$~V output referenced to SoM board ground. Figure~\ref{fig:spectrum} shows the output spectra in a band from $2$~Hz to $750$~MHz with a $10$~Hz resolution bandwidth.  The bandwidth is limited by the minimum value read by the spectrum analyzer, and the maximum bandwidth of the probe tool.  The figure shows that only noise present on the ground is discernibly measured on the channel output, with the maximum contributions reaching only $-70$~dBm.  The test setup is likely limited by the lack of shielding around the PCB, and the switched-mode power supplies used to power the board which could be potentially improved by using a low-noise, linear power supply and providing a grounded chassis for the SoM.

\subsection{Transient Behavior}

\begin{figure}[h]
\centering
\includegraphics[width=\linewidth]{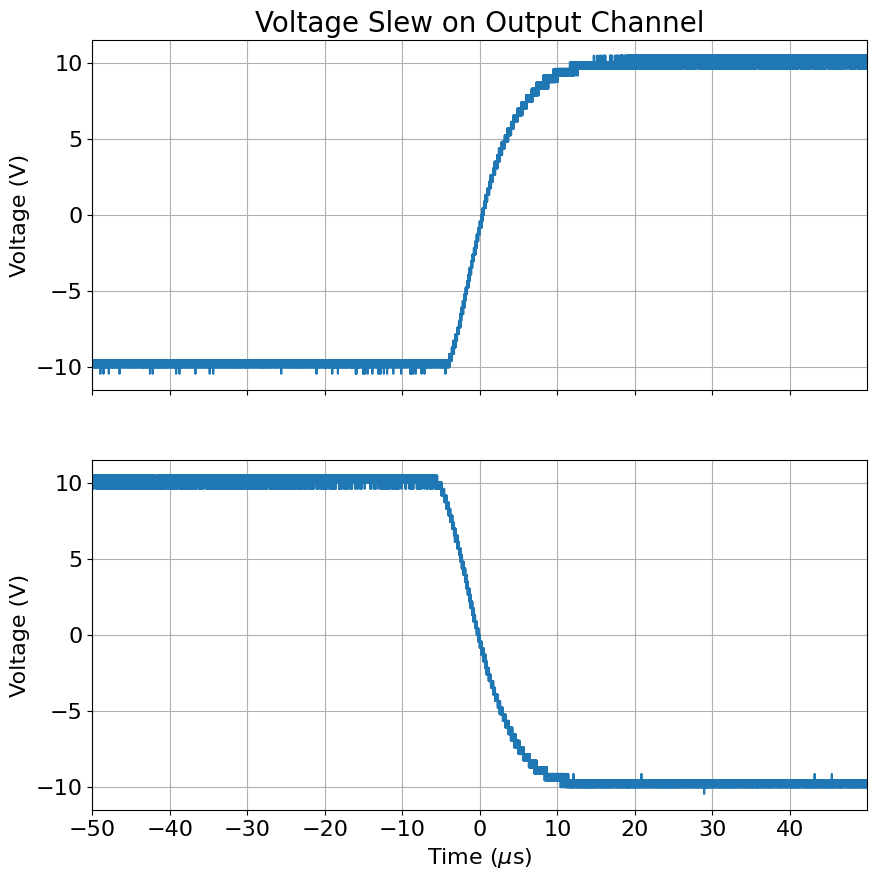}
\caption{Positive (top) and negative (bottom) full range voltage transition ($-10$~V to $10$~V) on DAC output channel, corresponding to an average slew rate of $1.76$~V/$\mu$s.}
\label{fig:slew}
\end{figure}

To benchmark the performance of the transient response of the output voltage channels, a measurement of a voltage transition was taken by a direct update of the the output control register from $0000h$ to FFFF$h$ corresponding to a shift directly from $-10$~V to $+10$~V.  The registers were written directly via a UART update handed off to the SPI module on FPGA which subsequently writes the values to the DAC output register, configured to update the output voltage immediately.  Figure~\ref{fig:slew} shows the waveform for a positive and negative transition, from $-10$~V to $+10$~V and $+10$~V to $-10$~V respectively.  The slew rate for transitions across all measured channels averages $1.76$~V/$\mu$s, lower than the specified slew rate on the DAC81416 datasheet for this voltage range of $4$~V/$\mu$s. This is a result of including the unipolar RC filter stage on the output of the SoM, which modestly restricts the update bandwidth.  An adjacent DAC channel with a fixed output voltage of $+1$~V was monitored using the high-precision oscilloscope during the voltage transitions, and no change was measured above the normal output stability (measurement presented in \ref{sec:measurement_1}).  This reinforces the expectation of the designed channel cross-talk mitigation measures.

\section{Outlook}
\label{sec:outlook}
The DAC ICs are independently feature rich, and the Spartan-7 FPGA provides an open environment for development of custom control logic.  The design presented in the foregoing is a baseline platform upon which continuous feature development can occur, leveraging the existing hardware, and expanding or adapting as the requirements in various experimental systems vary or scale.  Presented here is a non-exhaustive list of possible features and improvements which may be of use and could be employed in future iterations of the SoM.

% \color{red} Cost reduced version? \normalcolor

%\begin{enumerate}
%    \item Multiple input triggers X\\
%    \item Overtemp sensing, ALMOUT set/autoshutdown \\
%    \item Streaming mode register X\\
%    \item sync./async. update modes, using LDAC trigger for sync. ops X \\
%    \item CRC on SPI (might be useful eventually) X
%\end{enumerate}

The prototype control build makes no use of reading DAC registers, however several status registers are available on the DAC81416 which may be used when integrating the SoM, particularly in systems with multiple devices deployed.  The DAC81416 provides read registers for over-temperature sensing, alarms and auto-shutdown if required, as well as the ability to read back the content of the output data registers for verification of valid output data etc.  

For the prototype, a single SMA connectorized digital logic pin was included to provide a convenient external trigger input for setting the DAC configuration, or other timing critical operations.  There are however, many spare IO pins on the FPGA which may be employed for multiple parallel triggers or external switches which may be implemented in future hardware revisions.  For this version of the board, a USB-to-UART adapter cable which handles translation from the PC running the controller to the FPGA UART interface is used (described in section \ref{sec:digital_design}). However, in a system where UART serial is chosen as the means to transfer data, future revisions could include USB-to-UART communication hardware on board to simplify modular integration and usability.  Alternative communication protocols are also potentially desirable, particularly where the target system scales in complexity, such as I$^2$C or SPI for module-to-module communication.  In addition to the expanded FPGA design to facilitate alternate communication protocols, appropriate hardware can be included such as RJ45 or SFP.  This may be beneficial to facilitate higher bandwidth data streaming and synchronization across multiple modules, or to a primary control system such as the Sinara/ARTIQ framework.  Due to the ubiquity of the ARTIQ control framework and its use at the Duke Quantum Center, software integration for this control architecture is currently in development.

The prototype also uses asynchronous mode to set the DAC output immediately after a register write.  The DAC81416 has synchronous mode which can be exploited for direct, real-time and controlled latency of triggered output voltage settings in time-critical operations.  In synchronous mode, the DAC data registers are written asynchronously, but the output is not set until a dedicated trigger is set.  The DAC81416 has the capability of ``Streaming Mode Operation'' which may be useful to implement.  The standard method for updating the DAC channel registers require a large amount of data to be transferred to the device via SPI i.e. address register for each DAC channel and setup commands.  If Streaming Mode is employed, the DAC data registers can be written without having to provide each data register with an instruction command, shifting data continuously through the DAC registers, incrementing the address as long as the chip select line is asserted \cite{ti_dac81416}.  This may be particularly useful for high-speed ion-transport operations in real-time control scenarios. 

The DAC81416 also has the capability of operating in a daisy-chain, whereby the SPI communication can be concatenated across multiple devices, useful for reducing the number of serial interface lines.  Making use of this feature may be desirable as channel density scaling is required, expanding the number of DAC devices operated from the distributed control logic on module.  If deployed in a system where a very stable, accurate voltage reference is available, there is also the option to adapt the board to accommodate connection to an external voltage reference for the DACs, rather than relying on the internal voltage reference.

The connector footprint for the analog outputs in this prototype version accepts a 37-pin micro-D Sub type connector.  This was chosen purely for convenience of longer term testing and deployment in one of the ion-trap systems at the Duke Quantum Center.  This choice for the DAC SoM is arbitrary however, and future iterations of the board may be trivially branched or updated to accommodate a different connector standard if desired.  Another choice of convenience for this output connector was made, in that no explicit ground connection is made on the connector.  Once again, if this is a desired feature, a board revision to include it is relatively trivial, particularly with the current connector type having several spare channels available.

\section*{Data Availability}
Given the open development nature of the platform, a non-exhaustive list of potential improvements and developments, both in hardware and software have been discussed.  The authors invite testing, adoption and contribution to the version controlled project, publicly available on Gitlab at \url{https://gitlab.com/Coogani/vanguard-dac-som}.
%\IEEEpubidadjcol

% \section*{Acknowledgments}

% This material is based upon work supported by the U.S. Department of Energy, Office of Science, National Quantum Information Science Research Centers, Quantum Systems Accelerator (Award No. DE-SCL0000121). Additional support is acknowledged from the National Science Foundation: Software-Tailored Architecture for Quantum Co-Design (STAQ) Award (PHY-2325080) and QLCI: Center for Robust Quantum Simulation (OMA-2120757).

\section*{Author Contributions}
M.G.P. conceived of the project, designed and assembled the hardware, supervised implementation of the control architecture, and carried out experiments.
M.M.K. created and verified the digital design, wrote the UART control software, and carried out experiments.
C.N. supervised experiments, provided resources for prototyping and measurements, and acquired funding. 
All authors contributed to writing and reviewing the manuscript.

\bibliographystyle{IEEEtranN}
\bibliography{bibliography}

\end{document}